\documentclass[sigconf, nonacm]{acmart}
\newcommand\vldbdoi{XX.XX/XXX.XX}

\newcommand\vldbavailabilityurl{http://vldb.org/pvldb/format_vol14.html}
 
\usepackage{xcolor}
\usepackage{appendix}
\usepackage{xcolor}
\usepackage{xpatch}
\begin{document}
\title{IBM  Employee Attrition Analysis}

\author{Shenghuan Yang }
\affiliation{%
  \institution{Jiangxi University of Finance and Economics }
}

\author{Md Tariqul Islam}

\affiliation{%
  \institution{ Syracuse University}
}


\begin{abstract}
\bf In this paper, we analyzed the dataset IBM Employee Attrition to find the main reasons why employees choose to resign. Firstly, we utilized the correlation matrix to see some features that were not significantly correlated with other attributes and removed them from our dataset. Secondly, we selected important features by exploiting Random Forest, finding monthlyincome, age, and the number of companies worked significantly impacted employee attrition. Next, we also classified people into two clusters by using K-means Clustering.Finally, we performed binary logistic regression quantitative analysis: the attrition of people who traveled frequently was 2.4 times higher than that of people who rarely traveled. And we also found that employees who work in Human Resource have a higher tendency to leave.

\end{abstract}

\maketitle




\section{Introduction}

Employee attrition is defined as the natural process by which employees leave the workforce – for example, through resignation for personal reasons or retirement – and are not immediately replaced{\color{blue}{\cite{{ref5}}}}. Employee turnover is regarded as
the key issue for all organizations these days, because of its
adverse effects on workplace productivity, and accomplishing
organizational objectives on time {\color{blue}{\cite{{ref6}}}}.  In order for an organization to continually have a higher competitive advantage over its competition, it should make it a duty to minimize employee
attrition Therefore, for the better development of corporation, it is essential for the leader of companies to know the main reasons why their employees choose to leave the company,then take relevant measures to  improve their company's productivity, overall workflow and business performance.

\textbf{Objectives.} In this paper, we aim to select the main 
causes that contribute
to an employee’s decision to leave a company, and to be able to predict whether a particular employee
will leave the company  by utilizing machine learning models. 

\textbf{Contributions.} Following are the main contributions of this paper: 
\begin{itemize}
    \item We select the main factors affecting the employee attrition by using Random Forest, and classify which types of people are more likely to quit by utilizing the K-means Clustering 
    \item We represent a given reality in terms of a numerical value  to compare the employee attrition in different categories by utilizing quantitative analysis.


\end{itemize}
The rest of the paper is described below. We present
some related work in Section 2. We interpret data and introduce methodologies in Section 3. We process the original data set and remove some  variables which  are not very correlated with other features in section 4. We implement our machine learning model in Section 5. Finally, we draw a conclusion in section 6.

\section{Related Work}

Alao et al.{\color{blue}{\cite{{ref111}}}} used  decision tree models and rule-sets to develop a predictive model
 that was used to predict new cases of employee attrition. Alduayj et al.{\color{blue}{\cite{{ref222}}}}utilized support victor machine
(SVM) with several kernel functions, random forest and Knearest neighbour (KNN) to predict employee attrition based on their
features and found quadratic SVM scored the highest results. Frye et al.{\color{blue}{\cite{{ref333}}}}applied Principal Component Analysis and classification methods K-Nearest Neighbors and
 Random Forest, finding that Logistic Regression predicts employee quits with the highest accuracy. Yadavet et al.{\color{blue}{\cite{{ref444}}}} provided a framework for predicting the employee churn by analyzing the
employee’s precise behaviors and attributes using classification techniques. Srivastava et al.{\color{blue}{\cite{{ref555}}}}provided a framework for predicting the employee churn by analyzing the
employee’s precise behaviors and attributes using classification
techniques. El-Rayes et al.{\color{blue}{\cite{{ref666}}}} presented a framework
for predicting the employee attrition with respect to voluntary termination employing predictive analytics. Setiawan et al.{\color{blue}{\cite{{ref777}}}} found that eleven variables that have a significant impact on employee attrition.

\section{Data and Methodology}
Employee attrition is the internal data of the company, which is difficult to obtain, and some data has a certain degree of confidentiality, therefore our paper used the data set disclosed by kaggle. The sample size of the data set is 1471, there are 34 feature variables,mainly divided into three types of variables: personal basic information, work experience, attendance rate.This paper explored the relationship between employee’s  characteristics and employee attrition, found Whether characteristics have a great influence on employee attrition. In addition, we used machine learning algorithms to select important features that influenced the employee attrition, and predicted the it. In this paper, we exploited three machine learning algorithms: Decision Tree, and Logistic Regression and k-means clustering.

\subsection{Random Forest}

\begin{figure}[H]
  \centering
  \includegraphics[width=\linewidth]{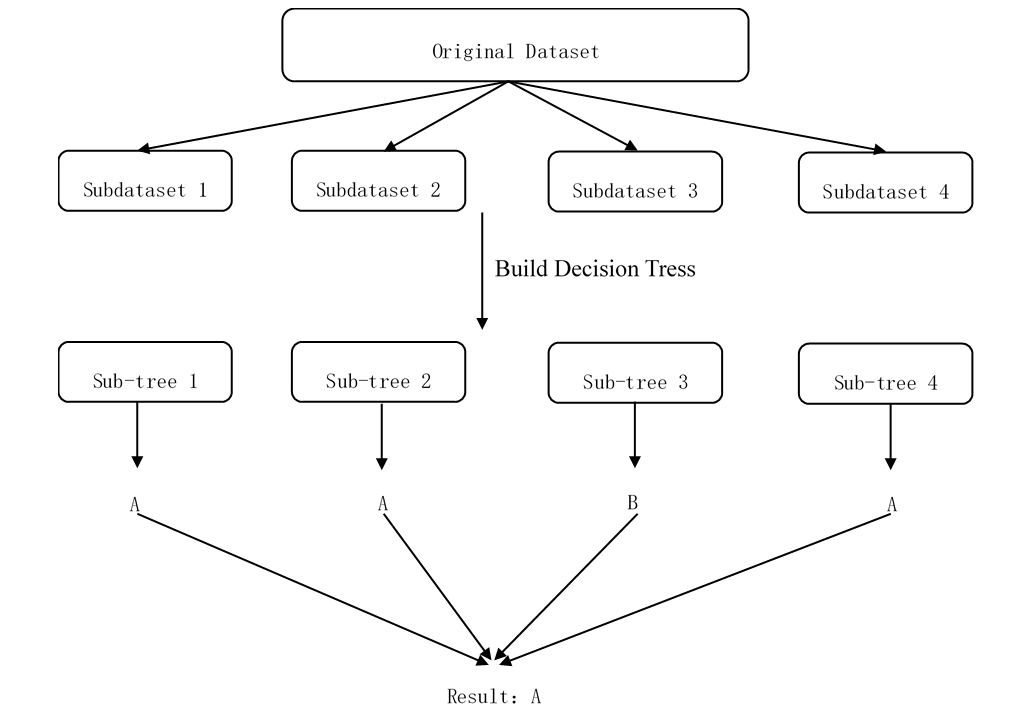}
  \caption{Decision-making process of random forest }
 \label{random forest}
\end{figure}

Random forest is an  learning method used for classification, regression and other tasks, which combines multiple decision trees to select the best result. Random forest corrects the habit of decision trees that rely too much on the training set and improves the accuracy of the model.First, there are randomly selected sub-data sets that are replaced from the original data{\color{blue}{\cite{{ref7}}}}. The elements between the sub-datasets may have the same elements, and the content of each sub-data set is 
1
 different. Then use the sub-dataset to construct the sub-decision tree, each decision tree will output a result, you can vote through the output result of the sub-decision tree, and finally get the output result. As shown in Figure {\color{blue}\ref{random forest}}, the data set extracts 4 sub-datasets to construct 4 sub-decision trees, 3 trees voted as A, one sub-decision tree voted as B, and the final output is A.

Random forest is one of the most popular machine learning algorithms. Random forest provides a unique combination of prediction accuracy and model
interpretability among popular machine learning methods. The random sampling
and ensemble strategies utilized in RF enable it to achieve accurate predictions as well as better generalizations {\color{blue}{\cite{{ref8}}}}. In addition, it has a strong explanatory nature. It can directly calculate the importance of each variable. In other words, it is easy to calculate how much each variable contributes to decision-making.
In our research, we build Random Forest model based on Employee Attrition Features.There are 34 employee attributes in the data set, we select randomly k(k<34)  employee attributes to build a decision tree, and create 100 random sub-samples of our dataset with replacement. Each node of each decision tree will be split according to predictor variables so that child nodes are more “pure” (i.e., Gender) in terms of the outcome variable. If the feature of the next child node has appeared in the parent node, the second child node will stop splitting. In the end, each decision tree outputs a result, and the result with the most votes is the final result.


\subsection{Binary Logistic Regression}

Logistic regression analysis can be used to estimate the probability of an event, and it can also analyze the influencing factors of the problem. In medical research, Logistic regression is often used to analyze the risk factors of a certain disease{\color{blue}{\cite{{ref99}}}}. Such as analyzing whether age, smoking, drinking, and diet are risk factors for diabetes. In questionnaire research, Logistic regression is often used to analyze non-scale questions, such as taking the basic background information of the sample as X and purchasing intention as Y to analyze whether gender, age, and  families' socioeconomic conditions affect purchase intention. Among them, binary logistic regression analysis is used most frequently.

\begin{equation}
 logit_(p)= \ln\dfrac {{p(y=1)}}{1-p(y=1)}={\beta}_0+{\beta_1}{x_1}+...+{\beta_n}{x_n}
\end{equation}

Here, employee attrition as the dependent variable y, 0 represents No, 1 represents Yes. Gender, EducationLevel, OverTime, JobLevel and so on as  predictor variables, n=34, ${x_1}$,${x_2}$,..., ${x_{34}}$. ${\beta_0}$, ${\beta_1}$,..., ${\beta_n}$ are coefficients, which represent the influence of the variable on the dependent variable. Whether it is positive or negative, it needs to be explained in conjunction with the corresponding regression coefficient value. If the regression coefficient value is greater than 0, it means a positive influence; otherwise, it means a negative influence for the dependent variable.


\subsection{k-means Clustering }

 The K-means Clustering algorithm is the most commonly used clustering algorithm. The main idea is: given K values and K initial cluster centers, assign each point  to the nearest cluster center, after all the points are allocated, the center point of the cluster is recalculated based on all the points in the cluster (take the average value), updates the center point of the cluster Steps until the center point of the cluster changes very little {\color{blue}{\cite{{ref10}}}}.
 The objective function is:
 \begin{equation}
   J=\sum_{j=1}^{k}\sum_{i=1}^{n}{||x^{(j)}_{i}-c_{j}||^{2}}   
 \end{equation}
Where $||x^{(j)}_{i}-c_{j}||$ is the Euclidean distance between $x^{(j)}_{i}$ and $c_{j}$, n is the number of data points in $j^{th}$ cluster, k is the number of cluster centers.
Algorithmic steps for k-means clustering:
$X = {x_{1},x_{2},x_{3},...,x_{n}}$ is the set of data points, and $c={c_{1},c_{2},...,c_{k}}$ is the set of centers.
\begin{enumerate}
    \item Randomly select 'k' cluster centers
    \item Calculate the distance between each data point and cluster centers
    \item Assign the data point to the cluster center whose distance from the cluster center is minimum of all the cluster centers
    \item Recalculate the new cluster center using:

\begin{equation}
       c_{j}=\dfrac{1}{n}\sum_{i=1}^{n}x_{i}
\end{equation}   

     \item Recalculate the distance between each data point and new obtained cluster centers
     \item  If no data point was reassigned then stop, otherwise repeat from step (3)

\end{enumerate}

\begin{figure}[H]
  \centering
  \includegraphics[width=\linewidth]{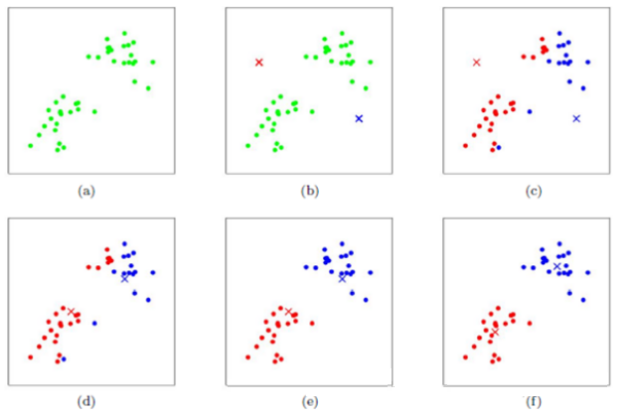}
  \caption{ Clustering  process of K-means Clustering }
 \label{K}
\end{figure}

The process of K-means Clustering is shown in Figure \ref{K}.
 
\section{data process}
In our studies, there are 34 variables, however, some features just have one data level that do not make sense for our research, such as 
EmployeeCount, Over18 and StandardHours, and employee number does not have meaning in analyzing resulting so we also deleted these features. In addition, some features are not very correlated with other attributes, we need to remove them from our dataset to improve computing efficiency. We built correlation matrix, which is a table showing correlation coefficients between variables. Each cell in the table shows the correlation between two variables.
 \begin{figure}[H]
  \centering
  \includegraphics[width=\linewidth]{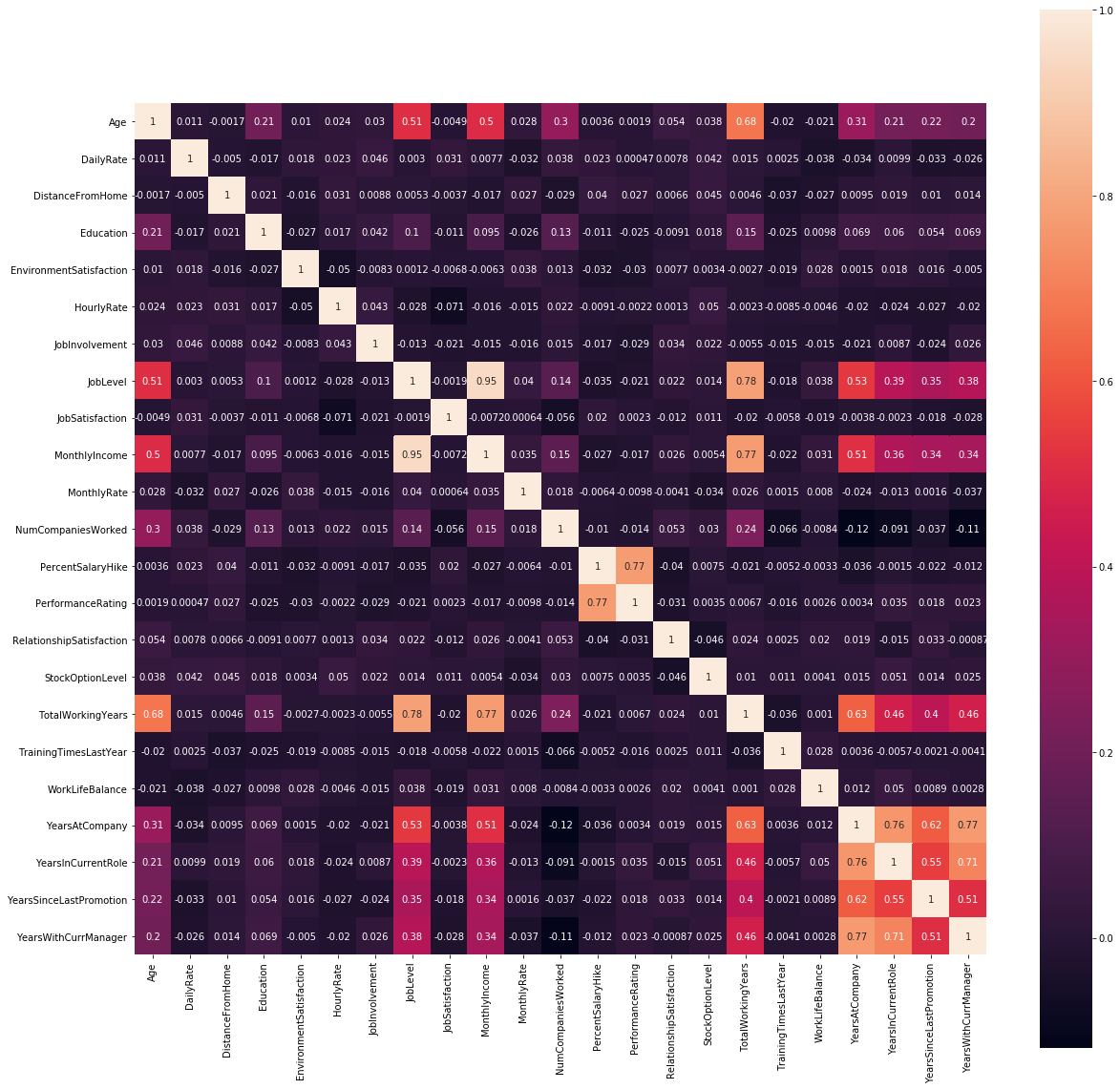}
  \caption{Correlation Matrix}
 \label{correlation}
\end{figure}
 
  We can see from Figure {\color{blue} \ref{correlation}}, daily rate, hourly rate and monthly rate are barely correlated with other attributes, while Monthly Income and  job level (0.95), Job level and total working years(0.78), Monthly Income and total working years(0.77) are highly correlated. Therefore we delete  daily rate, hourly rate and monthly rate from dataset.

\section{Model Construction}
In this part, we build machine learning models to select important features that influence the employee attrition and  classify the features to  help us understand the main reason why people left the IBM company.And we will construct the model to predict the employee attrition according to the given data.
\subsection{Feature Selection}
We all may have faced this problem of identifying the related features from a set of data and removing the irrelevant or less important features with do not contribute much to our target variable in order to achieve better accuracy for our model and.Therefore, it is better for us to select important features that influence the employee attrition, and ignore the variables that are not significant in explaining such behavior.

\begin{figure}[H]
  \centering
  \includegraphics[width=\linewidth]{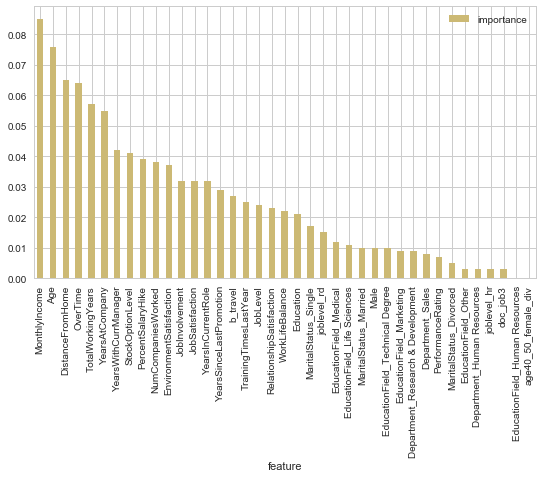}
  \caption{ Important Features }
 \label{feature}
\end{figure}

 From  Figure {\color{blue} \ref{feature}}, Monthly Income, Age, DisatnceFromHome are top 3 important features to indicate whether the employee has a tendency to leave, while marital status married, and female aged 40-50 are less likely to go. Indeed, income is the main cause that people choose to leave the company, which is highly related to people's life quality. More affluent people have more disposable
income and can more easily afford expensive service (such as medical care) and a healthy lifestyle—benefits.So people want to have more money that they intend to leave the current company and find other companies that offer high salaries. And also young people are more likely to want to try different jobs and finally find a job that suits them when they just graduated from universities. Whereas older people prefer a stable life because they already have own families and children, so it is not easy to leave the job. In addition, the distance between the company and home is also the main reason why people quit the current job, they would waste much time commuting. There's mass of social science and public health research on the negative effects of commuting on personal and societal well-being. Longer commutes are linked with increased rates of obesity, high cholesterol, high blood pressure, back and neck pain, divorce, depression and death.

Here, we need to train and test dataset to predict the employee attrition according to the features. We splited the data set into two sets: a training set and a testing set (80\% for training, and 20\% for testing). To better reflect the real situation and avoid the influence of extreme values, we  predicted the employee attrition  for 100 times by utilizing RandomForest Classifier, and recorded the average  accuracy of the test set.  Table {\color{blue} \ref{accuracy}} shows the average accuracy is 0.84561, which means our model fits well.

\begin{table}[hb]
\setlength\tabcolsep{3.5pt}
\caption{Prediction Accuracy of Random Forest }
 \label{accuracy}
  \setlength{\tabcolsep}{9mm}{
  \begin{tabular}{cc}
    \toprule
    Times & Accuracy \\
    \midrule
  1& 0.85714 \\
2&0.83673\\
3&0.84353\\
4&0.84693\\
...&...\\
100& 0.85374\\
Average Accuracy&0.84561\\

 \bottomrule
\end{tabular}
}
\end{table}
 
\subsection{Classification of people}
To distinguish which types of people are more likely to resign, we use K-means clustering to divide the dataset into two categories. The first type is prone to leave, and the other is less likely to quit. 

\begin{table}[hb]
\setlength\tabcolsep{3.5pt}
\caption{Clustering}
 \label{clustering}
{
  \begin{tabular}{ccc}
    \toprule

  Cluster Type	&	0	&	1	\\ 
  \midrule
Age	&	44.215152	&	34.813158	\\
Attrition	&	0.1	&	0.178947	\\
DistanceFromHome	&	9.072727	&	9.227193	\\
Education	&	3.039394	&	2.876316	\\
EnvironmentSatisfaction	&	2.693939	&	2.729825	\\
JobInvolvement	&	2.690909	&	2.741228	\\
JobLevel	&	3.684848	&	1.594737	\\
JobSatisfaction	&	2.709091	&	2.734211	\\
MonthlyIncome	&	14060.49394	&	4315.215789	\\
NumCompaniesWorked	&	3.342424	&	2.505263	\\
OverTime	&	0.290909	&	0.280702	\\
...&...&...\\
...&...&...\\

PercentSalaryHike	&	15.066667	&	15.250877	\\
PerformanceRating	&	3.148485	&	3.155263	\\
RelationshipSatisfaction	&	2.781818	&	2.692105	\\
StockOptionLevel	&	0.80303	&	0.791228	\\
TotalWorkingYears	&	21.072727	&	8.444737	\\

 \bottomrule
\end{tabular}
}
\end{table}

We can see from Table {\color{blue}\ref{clustering}} (complete form in appendix table {\color{blue}\ref{coK-mean}}),cluster set 0  represents low attrition, cluster set 1 means high attrition, and older people, high job level, high job satisfaction,high monthly income, more number of companies worked and so forth are less likely to leave.These finding are in line with people's behavior in the real world and previous accounts in feature selection of RandomForest. In last section, we illustrated the relationship between age, monthly income, distance from the company and home, this section, we also find that the number of companies employees worked are related to the probability of  leaving the 
corporation. People who have worked in 3-4 companies are less likely to quit because by this time, they have roughly found the direction of employment, and people who have more than four companies indicate that they are unstable and often change jobs. In addition, people who are in the higher the job level, enjoy the higher salary and respect , so they are less likely to leave. While those employees in lower-level positions are often not satisfied with the status quo and want to  seek better job chance. What is more, job satisfaction is also the main cause influencing the employee attrition rate.Intention
to stay on the job is clearly correlated with job satisfaction in such aspects as educational system and environment, income and welfare, leadership and administration.{\color{blue}{\cite{{ref4}}}}.



\subsection{Logistics Regression}
In this section, we exploited the binary logistics regression to predict the relationship between predictors (employee characteristics) and a predicted variable (employee attrition) where the dependent variable is binary (NO:0, YES:1). And we will  compare the differences between each category to help us understand which type of person is more likely to quit (complete regression table can be seen in appendix table {\color{blue}\ref{COlogistics}}). 

\begin{table}[hb]
\setlength\tabcolsep{3.5pt}
\caption{Regression Result}
 \label{regression}
 \setlength{\tabcolsep}{5mm}{
  \begin{tabular}{ccc}
     \toprule

\quad		&	OR(95\% CI)				&	P-value	\\
\cline{2-3}
Travel$\_$Rarely	&	1	&		\\
Non\-Travel	&	0.361	&		\\
Travel\_Frequently	&	2.411	&	<0.05	\\
\cline{2-3}
Male	&	1	&		\\
Female	&	0.659	&	<0.05	\\
\cline{2-3}
Sales Representative	&	1	&		\\
Healthcare	&	0.160	&		\\
Human Resource	&	4.060	&		\\
Laboratory	&	0.556	&		\\
Manager	&	0.347	&		\\
Manufacturing	&	0.200	&		\\
Research	&	0.200	&		\\
Sales Executive	&	0.484	&	<0.05	\\
\cline{2-3}
Single	&	1	&		\\
Divorced	&	0.304	&		\\
Married	&	0.427	&	<0.05	\\
\cline{2-3}
OverTime	&	0.138	&	<0.05	\\

 \bottomrule
\end{tabular}
}
\end{table}

From Table  {\color{blue}\ref{regression}}, we can see, taking people who rarely travel as the standard 1, employees who travel frequently are 0.361 times more than those to leave the company whereas people who never travel are more than twice. And Women are 0.659 times more than men to go. As for JobRole, taking Sales Representative as the standard 1, employee who who work about Research Science, Laboratory, Manufacturing, Healthcare are not likely to quit, while HumanResource Department has high employee attrition. Then single people are more likely to leave their jobs than married and divorced people.Finally,not surprised, people who usually work overtime would like to leave the company.
To evaluate the logistics regression model, we need to test the accuracy of model in Python, the test result is 0.8843, which means our model fitting is well.

   


\section{conclusion}
According the above model results, we can know that our finding are in line with people's behavior in the real world and previous studies other scholar did. 
We utilized Random Forest and K-means Clustering to select important features that had obvious impact on the employee attrition. Firstly, according to Random Forest results, monthlyincome, age, the number of companies worked are the main reasons why people choose to resign. Then we found older people, high job level, high job satisfaction,high monthly income, more number of companies worked, these kinds of people are not likely to go based on the clustering result of K-means Clustering. However, different people have various intention, we need to do further and detailed research to find qualitative findings by using qualitative analysis. So we exploited the binary logistics regression to compare the difference between people. Our study found that females' attrition was 0.659 times than that of males, married and divorced people were 0.427 and 0.304 times than people who were single, respectively. Besides, the attrition of people who traveled frequently was 2.4 times higher than that of people who rarely traveled. And we also found that employees who work in Human Resource have a higher tendency to leave. Finally, there are other interesting findings in our study:in terms of number of companies worked, people who worked in 2 - 4 companies are less likely to leave, the female attrition rate is less than male after working for six companies , and people who earned Doctor's Degree are almost always having the lowest attrition rate.

To evaluate the model performance, we trained and tested the dataset to predict the employee attrition, split it into two parts(80\% for training, 20\% for testing ), and recorded the test set's accuracy. Random Forest and Logistics Regression accuracy were 0.8456 and 0.8843, respectively, which meant  Logistics Regression fitted better and was more suitable for prediction in our dataset.

We also want to make some suggestions to the company through this research, hoping that they will care more about their employees and improve their job satisfaction. Simultaneously, they must pay more attention to human resources employees because they have very low job satisfaction. Besides, the company should allow employees to have enough time to rest and spend time with their families.There is a general belief that employees who take regular breaks are more productive.

\begin{appendix}

\begin{table*}[t]
  \caption{Clustering Result}
  \label{coK-mean}
  \setlength{\tabcolsep}{3mm}{
  \begin{tabular}{ccc}
    \toprule

   Cluster Type	&	0	&	1	\\
   \midrule
Age	&	44.215152	&	34.813158	\\
Attrition	&	0.1	&	0.178947	\\
DistanceFromHome	&	9.072727	&	9.227193	\\
Education	&	3.039394	&	2.876316	\\
EnvironmentSatisfaction	&	2.693939	&	2.729825	\\
JobInvolvement	&	2.690909	&	2.741228	\\
JobLevel	&	3.684848	&	1.594737	\\
JobSatisfaction	&	2.709091	&	2.734211	\\
MonthlyIncome	&	14060.49394	&	4315.215789	\\
NumCompaniesWorked	&	3.342424	&	2.505263	\\
OverTime	&	0.290909	&	0.280702	\\
PercentSalaryHike	&	15.066667	&	15.250877	\\
PerformanceRating	&	3.148485	&	3.155263	\\
RelationshipSatisfaction	&	2.781818	&	2.692105	\\
StockOptionLevel	&	0.80303	&	0.791228	\\
TotalWorkingYears	&	21.072727	&	8.444737	\\
TrainingTimesLastYear	&	2.79697	&	2.8	\\
WorkLifeBalance	&	2.781818	&	2.755263	\\
YearsAtCompany	&	11.927273	&	5.584211	\\
YearsInCurrentRole	&	6.133333	&	3.67807	\\
YearsSinceLastPromotion	&	4.109091	&	1.631579	\\
YearsWithCurrManager	&	5.821212	&	3.631579	\\
Male	&	0.581818	&	0.605263	\\
BusinessTravelNon-Travel	&	0.087879	&	0.10614	\\
BusinessTravelTravelFrequently	&	0.187879	&	0.188596	\\
BusinessTravelTravelRarely	&	0.724242	&	0.705263	\\
DepartmentHuman Resources	&	0.045455	&	0.042105	\\
DepartmentResearch $\&$ Development	&	0.642424	&	0.657018	\\
DepartmentSales	&	0.312121	&	0.300877	\\
EducationFieldHuman Resources	&	0.021212	&	0.017544	\\
EducationFieldLife Sciences	&	0.39697	&	0.416667	\\
EducationFieldMarketing	&	0.130303	&	0.101754	\\
EducationFieldMedical	&	0.324242	&	0.313158	\\
EducationFieldOther	&	0.045455	&	0.058772	\\
EducationFieldTechnical Degree	&	0.081818	&	0.092105	\\
JobRoleHealthcare Representative	&	0.112121	&	0.082456	\\
JobRoleHuman Resources	&	0.012121	&	0.042105	\\
JobRoleLaboratory Technician	&	0	&	0.227193	\\
JobRoleManager	&	0.309091	&	0	\\
JobRoleManufacturing Director	&	0.121212	&	0.092105	\\
JobRoleResearch Director	&	0.242424	&	0	\\
JobRoleResearch Scientist	&	0.00303	&	0.255263	\\
JobRoleSales Executive	&	0.2	&	0.22807	\\
JobRoleSales Representative	&	0	&	0.072807	\\
MaritalStatusDivorced	&	0.251515	&	0.214035	\\
MaritalStatusMarried	&	0.506061	&	0.44386	\\
MaritalStatusSingle	&	0.242424	&	0.342105	\\

    \bottomrule
  \end{tabular}
  }
\end{table*}

\begin{table*}[t]
  \caption{Logistics Regression Result}
  \label{COlogistics}

\begin{tabular}{ccccccccc}
		
	\toprule

\multicolumn{7}{c}{}	&	\multicolumn{2}{c}{95\% C.I.for EXP(B)}			\\
\cline{8-9}
	&	B	&	S.E.	&	Wald	&	df	&	Sig.	&	Exp(B)	&	Lower	&	Upper	\\
	 \midrule
Age	&	-0.031	&	0.014	&	5.285	&	1	&	$0.02^{**}$	&	0.969	&	0.944	&	0.995	\\
TravelRarely	&		&		&	27.561	&	2	&	$0.00^{**}$	&		&		&		\\
NonTravel	&	-1.020	&	0.381	&	7.175	&	1	&	$0.02^{**}$	&	0.361	&	0.171	&	0.761	\\
TravelFrequently	&	0.880	&	0.211	&	17.323	&	1	&	$0.04^{**}$	&	2.411	&	1.593	&	3.649	\\
DailyRate	&	0.000	&	0.000	&	1.645	&	1	&	0.200 	&	1.000	&	0.999	&	1.000	\\
Sales Department
	&		&		&	0.010	&	2	&	0.995 	&		&		&		\\
HumanResource Department
	&	-18.223	&	10790.424	&	0.000	&	1	&	0.999 	&	0.000	&	0.000	&		\\
ResearchDepartment
	&	0.104	&	1.043	&	0.010	&	1	&	0.921 	&	1.109	&	0.144	&	8.566	\\
DistanceFromHome	&	0.046	&	0.011	&	17.837	&	1	&	$0.00^{**}$	&	1.047	&	1.025	&	1.069	\\
Education	&	0.007	&	0.088	&	0.006	&	1	&	0.937 	&	1.007	&	0.848	&	1.196	\\
Human Re	&		&		&	13.256	&	5	&	$0.02^{**}$	&		&		&		\\
Life Science	&	-0.189	&	0.823	&	0.053	&	1	&	0.818 	&	0.828	&	0.165	&	4.152	\\
Marketing	&	-0.926	&	0.304	&	9.279	&	1	&	$0.00^{**}$	&	0.396	&	0.218	&	0.719	\\
Medical	&	-0.525	&	0.394	&	1.777	&	1	&	0.182 	&	0.591	&	0.273	&	1.280	\\
Other	&	-1.036	&	0.312	&	11.029	&	1	&	$0.00^{**}$	&	0.355	&	0.193	&	0.654	\\
Technical	&	-0.983	&	0.475	&	4.276	&	1	&	$0.00^{**}$	&	0.374	&	0.147	&	0.950	\\
EmployeeNumber	&	0.000	&	0.000	&	0.888	&	1	&	0.346 	&	1.000	&	1.000	&	1.000	\\
EnvironmentSatisfaction	&	-0.431	&	0.083	&	26.988	&	1	&	$0.00^{**}$	&	0.650	&	0.553	&	0.765	\\
Gender(1)	&	-0.417	&	0.184	&	5.098	&	1	&	$0.00^{**}$	&	0.659	&	0.459	&	0.947	\\
HourlyRate	&	0.001	&	0.004	&	0.087	&	1	&	0.768 	&	1.001	&	0.993	&	1.010	\\
JobInvolvement	&	-0.528	&	0.123	&	18.417	&	1	&	$0.00^{**}$	&	0.590	&	0.463	&	0.751	\\
JobLevel	&	-0.128	&	0.317	&	0.162	&	1	&	0.688 	&	0.880	&	0.472	&	1.640	\\
Sales Representative	&		&		&	20.855	&	7	&	$0.00^{**}$	&		&		&		\\
Healthcare	&	-1.835	&	1.164	&	2.484	&	1	&	0.115 	&	0.160	&	0.016	&	1.563	\\
Human Resource	&	17.601	&	10790.424	&	0.000	&	1	&	0.999 	&	44066714.479	&	0.000	&		\\
Laboratory	&	-0.588	&	1.090	&	0.291	&	1	&	0.590 	&	0.556	&	0.066	&	4.702	\\
Manager	&	-1.058	&	0.985	&	1.154	&	1	&	0.283 	&	0.347	&	0.050	&	2.392	\\
Manufacturing	&	-1.609	&	1.162	&	1.919	&	1	&	0.166 	&	0.200	&	0.021	&	1.950	\\
Research	&	-1.611	&	1.097	&	2.156	&	1	&	0.142 	&	0.200	&	0.023	&	1.715	\\
Sales Executive	&	-0.726	&	0.385	&	3.555	&	1	&	0.059 	&	0.484	&	0.228	&	1.029	\\
JobSatisfaction	&	-0.414	&	0.081	&	25.891	&	1	&	$0.00^{**}$	&	0.661	&	0.563	&	0.775	\\
Single	&		&		&	14.274	&	2	&	$0.00^{**}$	&		&		&		\\
Divorced	&	-1.191	&	0.346	&	11.885	&	1	&	$0.00^{**}$	&	0.304	&	0.154	&	0.598	\\
Married	&	-0.852	&	0.251	&	11.525	&	1	&	$0.00^{**}$	&	0.427	&	0.261	&	0.698	\\
MonthlyIncome	&	0.000	&	0.000	&	0.292	&	1	&	0.589 	&	1.000	&	1.000	&	1.000	\\
MonthlyRate	&	0.000	&	0.000	&	0.138	&	1	&	0.710 	&	1.000	&	1.000	&	1.000	\\
NumCompaniesWorked	&	0.196	&	0.039	&	25.654	&	1	&	$0.00^{**}$	&	1.216	&	1.128	&	1.312	\\
OverTime(1)	&	-1.984	&	0.194	&	104.913	&	1	&	$0.00^{**}$	&	0.138	&	0.094	&	0.201	\\
PercentSalaryHike	&	-0.021	&	0.039	&	0.287	&	1	&	0.592 	&	0.979	&	0.907	&	1.058	\\
PerformanceRating	&	0.096	&	0.398	&	0.058	&	1	&	0.810 	&	1.100	&	0.504	&	2.402	\\
RelationshipSatisfaction	&	-0.270	&	0.083	&	10.591	&	1	&	$0.00^{**}$	&	0.763	&	0.649	&	0.898	\\
StockOptionLevel	&	-0.186	&	0.158	&	1.387	&	1	&	0.239 	&	0.830	&	0.609	&	1.131	\\
TotalWorkingYears	&	-0.057	&	0.029	&	3.772	&	1	&	0.052 	&	0.945	&	0.892	&	1.001	\\
TrainingTimesLastYear	&	-0.193	&	0.073	&	6.976	&	1	&	$0.00^{**}$	&	0.825	&	0.715	&	0.951	\\
WorkLifeBalance	&	-0.377	&	0.124	&	9.298	&	1	&	$0.00^{**}$	&	0.686	&	0.538	&	0.874	\\
YearsAtCompany	&	0.098	&	0.039	&	6.348	&	1	&	$0.00^{**}$	&	1.103	&	1.022	&	1.191	\\
YearsInCurrentRole	&	-0.148	&	0.045	&	10.701	&	1	&	$0.00^{**}$	&	0.862	&	0.789	&	0.942	\\
YearsSinceLastPromotion	&	0.175	&	0.042	&	16.934	&	1	&	$0.00^{**}$	&	1.191	&	1.096	&	1.294	\\
YearsWithCurrManager	&	-0.138	&	0.047	&	8.575	&	1	&	$0.00^{**}$	&	0.871	&	0.794	&	0.955	\\
Constant	&	9.308	&	1.383	&	45.275	&	1	&	$0.00^{**}$	&	11022.507	&		&		\\

  \bottomrule
**significant at p < 0.05	
	\end{tabular}

\end{table*}

\end{appendix}

\end{document}